\documentclass[a4paper]{article}
\usepackage{RR}
\RRNo{6396}
\usepackage{hyperref}
\usepackage{epsfig}
\usepackage{amssymb}

\RRdate{December 2007}

\RRauthor{J\'er\^ome Galtier, Orange Labs
}
\authorhead{Galtier}
\RRtitle{Tournament MAC with constant size congestion
window for WLAN}
\RRetitle{Tournament MAC with Constant Size Congestion
Window for WLAN}
\RRtitle{MAC en tournoi avec fen\`etre de congestion constante pour le WLAN}

\titlehead{Tournament MAC for WLAN}
\RRresume{
Dans le contexte des r\'eseaux radio distribu\'es, nous pr\'esentons
une approche g\'en\'eralis\'ee du protocole de controle d'acc\`es (MAC
pour {\it medium access control})) avec une fen\`etre de congestion
\`a taille fixe. Notre protocole est simple \`a analyser et
peut-\^etre utilis\'e dans beaucoup de contextes diff\'erents. Nous
donnons des argument math\'ematiques pour prouver que la performance
du protocol est optimale, dans le sens o\`u aucun protocole avec fen\`etre de congestion fixe ne peut l'am\'eliorer. Nous nous pla\c{c}ons
aussi dans le contexte WiFi/WiMAX, et nous montrons
exp\'erimentalement une r\'eduction du taux de collision entre 14\% et
21\% par rapport aux meilleures m\'ethodes connues. Nous montrons aussi l'am\'elioration obtenue en capaci\'e de
canal, et analysons les questions d'\'equit\'e pour ce protocole.}
\RRabstract{
In the context of radio distributed networks, we present a generalized
approach for the Medium Access Control (MAC) with fixed congestion
window. Our protocol is quite simple to analyze and can be used in a
lot of different situations.  We give mathematical evidence showing
that our performance is tight, in the sense that no protocol with
fixed congestion window can do better.  We also place ourselves in the
WiFi/WiMAX framework, and show experimental results enlightening
collision reduction of 14\% to 21\% compared to the best known other methods.
We show channel capacity improvement,
and fairness considerations.  }
\RRmotcle{R\'eseau radio local, WiFi, WiMAX, IEEE 802.11, efficacit\'e spectrale}
\RRkeyword{WLAN, WiFi, WiMAX, IEEE 802.11, spectral efficiency, radio channel}
\RRprojets{Mascotte}
\RRtheme{\THCom} 
\URSophia 

\newenvironment{proof}
   {\noindent \begin{rm}{\sc Proof.~}}
   {\hspace*{\fill}$\square$\par\end{rm}\vspace{\baselineskip}}

\begin{document}
\makeRR   

\def\N{\hbox{\rm I$\mskip-\thinmuskip$N\/}}
\def\R{\hbox{\rm I$\mskip-\thinmuskip$R\/}}

\bibliographystyle{plain}

\newtheorem{definition}{Definition}
\newtheorem{prob}{Problem}
\newtheorem{lem}{Lemma}
\newtheorem{prop}{Proposition}
\newtheorem{res}{Result}
\newtheorem{theoreme}{Theorem}
\newtheorem{cor}{Corollaire}


\section{Introduction}

Radio networks have received in the past years a growing interest for
their ability to offer relatively wide band radio
networking. Applications cover a large area of domains including
computer network wireless infrastructures, and high speed Internet
access for rural areas.

 In particular, in WiFi and partly WiMAX norms, the underlying
 mechanism\cite{IEEEnorm,IEEE16norm} (see also \cite{SLC06}) is a
 2-layer protocol whose first part relies on a derivative of the
 Binary Exponential Back-off protocol (BEB). The principle is that when
 a failure occurs, the transmission protocol delays the following
 retransmission by some penalty factor. The protocol uses a contention
 window (CW) mechanism to realize this back-off mechanism. Roughly
 speaking, the probability of trying an access to the channel is
 1/CW. When a transmission fails, the station increases $CW$ in order
 to be less demanding for futures accesses.

Already much research work has been done to model the $CW$
in\-crea\-se~/\-de\-crea\-se process. Strong simplifying
assumptions are at the basis of some models \cite{CCG00}, while some others
focus on an individual station while considering that the effect of
the others on the channel can be represented by an occupancy
probability $p_{occ}$ (see \cite{Bianchi00,WPL+02,Xiao03}), following an
earlier popular approach on CSMA \cite{KT75,BC88}.  

In fact, those studies show that on average, the contention window
mechanism draw the stations to access the channel with some probability
that converges to a value noted $p_{acc}(n)$ which depends on the number $n$ of
stations simultaneously willing to access the channel.  Therefore
several studies have shown that the optimal behavior
is when $p_{acc}(n)=O(1/n)$, and proposed some alternative mechanisms
to in\-crea\-se and de\-crea\-se the contention window in order to reach this
value. In \cite{BCG04}, the authors aim at guessing the total number 
of stations trying to emit in order to directly set the value $CW$.
In \cite{Galtier04a}, an optimization of the in\-crea\-se~/\-de\-crea\-se parameters
is done to converge to the optimal channel efficiency in terms of capacity.
In \cite{HRG+05} the authors use the observation of idle slots to deduce
the probable number of competing stations.

A different branch of CSMA protocols has been initiated by the
Hiperlan protocol \cite{Hiperlan}, a twin standard of 802.11a
developed in the same period. In this protocol, the contention phase
is bounded. The contention phase begins for each terminal by the
emission of a burst whose length follows a truncated geometric
distribution, and the terminal having the longest burst wins the right
to transmit. If several terminals have the same longest burst, this
results in a collision. A very similar protocol developed in the
context of sensor networks has been called Sift \cite{JBT03}.

Another related protocol is the {\it Contention Tree Algorithm}
\cite{Capetanakis79a,Capetanakis79b,JdJ00}, or CTA for short, also
called {\it Stack Algorithm} \cite{FFH+85,TM78}, which uses a
tree to solve contention problems. Although we also use a tree, our
protocol is completely different. This algorithm is basically based on
feedback, that is evidence that a collision occurred. In the radio
context, a feedback is expensive since it requires an acknowledgment
packet. On the contrary, we rely on {\it evidence of occupation}, which
is the fact that a silent terminal can detect that one or more
terminals are signaling their presence.

\begin{figure*}
\centerline{\psfig{figure=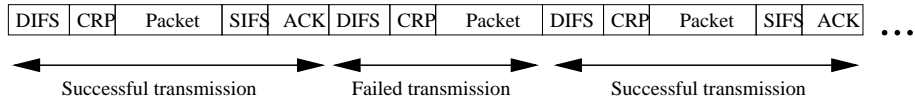,width=12cm}}
\caption{A view of the general frame structure.}
\label{fig:protocol}
\end{figure*}

In figure~\ref{fig:protocol} we show how our protocol works.  As in
the standard 802.11 approach, the transmission begins with a period of
sensing after the last emission (either an acknowledgment packet or a
failed one, for instance due to collision). After observing a
sufficiently long period with no emission (the DIFS period), the
system operates a contention resolution protocol (CRP) that is
supposed to select one station and only one. Then the packet is
transmitted. If it is correctly received, the receiving station emits
an ACK after a SIFS period. This is the end of a transmission period
and a new transmission period can begin. If no ACK is listened to, the
new transmission period begins immediately after the failed packet,
and the stations start the CRP mechanism just after the DIFS period
(recall that SIFS$<$DIFS).

How does the CRP work? In our protocol the time is subdivided into
time-slots that correspond to rounds of selection.  At the beginning,
each terminal emits a signal on the first time-slot with a certain
probability. When the station does not emit, it listens
to other signals, and, when it hears at least one other
signal, it withdraws itself from the selection. This process - called
{\em round} - is repeated $k$ times, where $k$ is a fixed integer, in
order to leave only one remaining station at the end with the highest
probability.  This method is used in \cite{AC05}.

The present article generalizes this method, and present a
mathematical framework to analyze its strengths and weaknesses.  As a
result, the improved method present a reduction of collision between
13.9\% and 21.1\%, resulting in a systematic gain of throughput.
Improving CONTI by a few percents, the gain to the original 802.11b
norm is as high as 31.4\%, achieving the best known throughput for
this family of protocols. Moreover, the new protocol
keeps excellent fairness characteristics, as indicates
the Jain index. 

Our experiments advocate that six mini-slots of selections (as in
\cite{AC05}) is indeed the correct number when the amount of contending
that this is indeed the correct number when the amount of contending
stations is often between 10 and 100. Anyway, our analysis can be very
easily extended to a different number than six mini-slots of selection,
so that by adding a sufficient (and provably optimal) number of
mini-slots, we can reduce the number of collisions to an arbitrarily
low level. However, this does not necessarily increase the throughput
since adding a mini-slot can be statistically more penalizing from the
throughput point of view than retransmitting a packet in case of collision.

The key to obtain those results is in the choice of the probabilities
with which a station will decide to keep silent or emit a signal in the CRP phase. Each
station takes into account the fact that signals were emitted or
not in the previous rounds to adapt its probability of emission.

In the following, we call try-bit and denote by $r(t)$ the fact that
at least one station has emitted a signal on the $t^{\mbox{\tiny th}}$
round.

\begin{figure}
\centerline{\psfig{figure=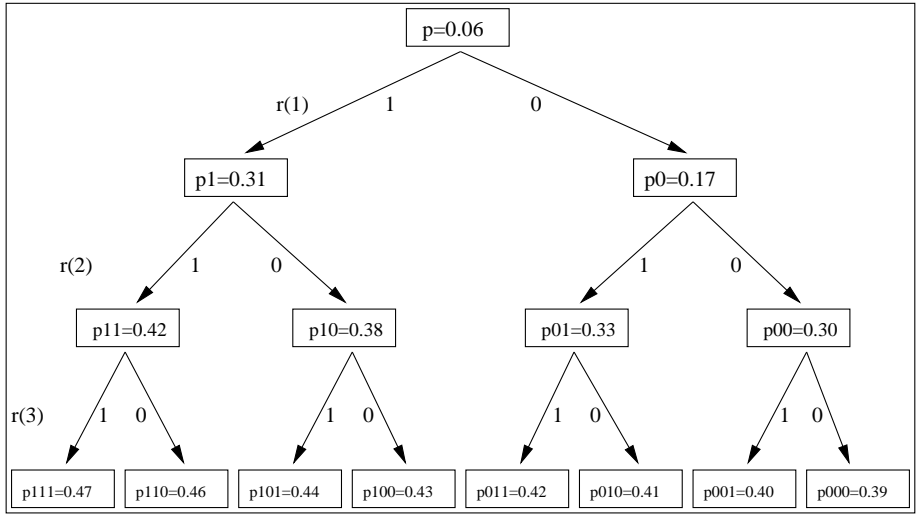,width=12cm}}
\caption{Choice of the values of $p$ in the course of the rounds of selection}
\label{fig:802.11c}
\end{figure}

\begin{figure*}
\centerline{\psfig{figure=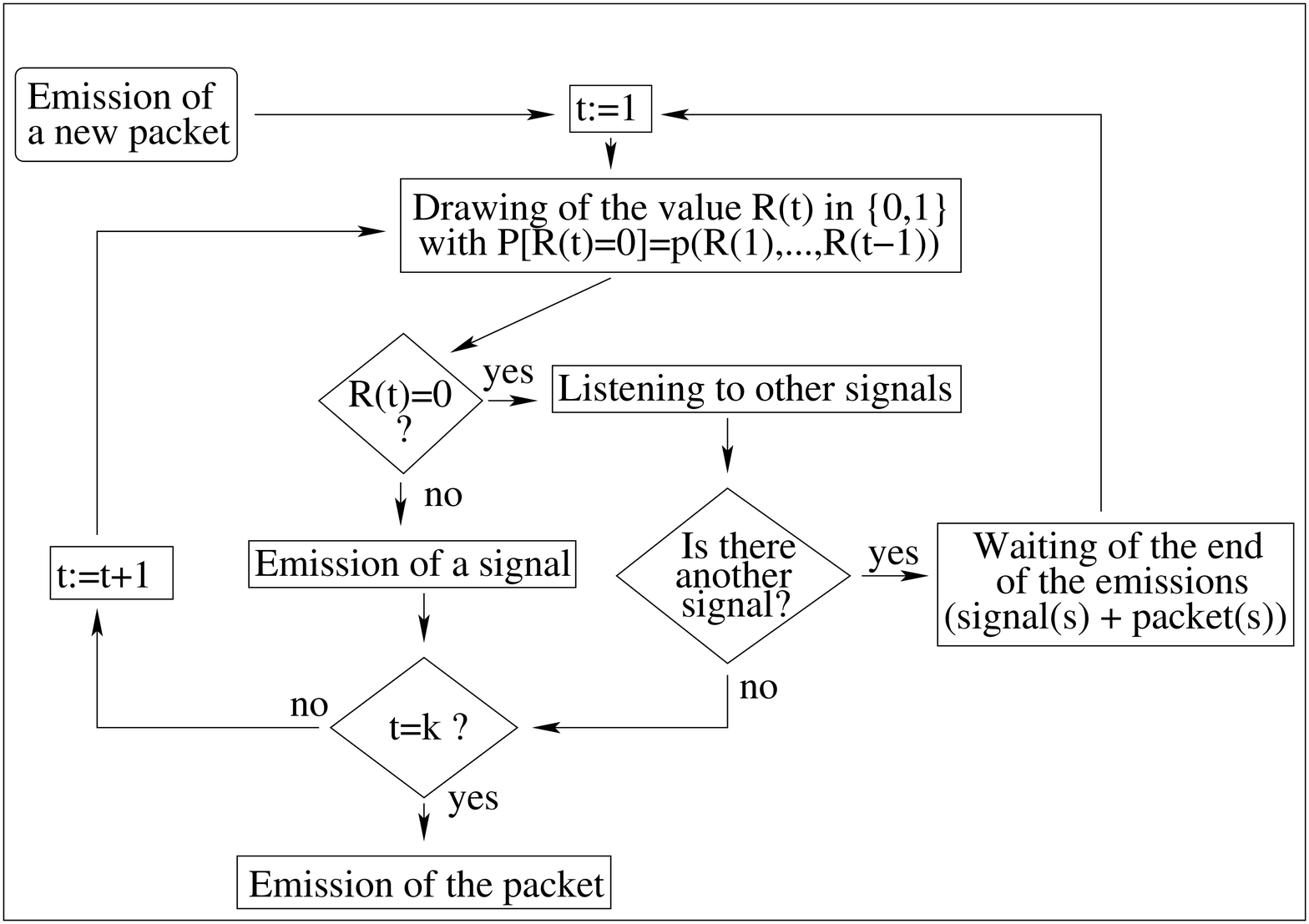,width=12cm}}
\caption{Global diagram for the rounds of selection in a terminal.}
\label{fig:map}
\end{figure*}

In figure~\ref{fig:802.11c}, we show how the choice of $p$ - the
probability that a station emits at a given round of selection -
evolves in the course of the rounds of selection and in function of
the previous try-bits chosen by the terminal. The first value (in the
figure, $p=0.06$) is unique for all the terminals. During the second
round, if the terminal has emitted a signal at the first round (which
we denote $r(1)=1$), the protocol chooses the left part of the tree,
and uses $p_1=0.31$ for the second round. If on the contrary the
terminal did not emit, and did not hear any signal of other
terminals ($r(1)=0$), then the protocol chooses the right part of the
tree and uses $p_0=0.17$ for the second round. If the
terminal did not emit and actually listens to a signal from another
station, it withdraws and leaves to other stations the right to send the
following packet.  As a result, the probability used in the second
round is necessarily $p_{r(1)}$. The realization of the second round
will determine the value of $r(2)$, and the third round will be
governed by the probability $p_{r(1)r(2)}$ in the tree of
figure~\ref{fig:802.11c}. We plot the whole process in
figure~\ref{fig:map}. Note that each station has a local try-bit
$R(t)$ at round $t$ which equals $r(t)$ as long as the station
is not eliminated.

We manage to find a tight approximation of the behavior of this
protocol when the number of rounds $k$ increases. More precisely, if we denote by $q_n$ the
probability that $n$ stations try to emit in the system, and if we
introduce the function
\begin{equation}\label{eq:deff}
 f(x)=\sum_{n\geq1}q_nx^n,
\end{equation}
then we can show that tuning the $p$ coefficients to this probability
space, the lowest possible rate of collision $1-\rho$ observed will be fairly approximated by
$$1-\rho\approx\frac{\left(\int_0^1\sqrt{f''(t)}dt\right)^2}{2^{k+1}}.$$

The article is organized as follows. A mathematical analysis in the
next section investigates analytically the optimization issues raised
by the problem of the choice of the $p$'s, and gives some tight bounds
for this question. The reader that desires to know the protocol
without the mathematics can skip this section.  A practical
implementation of the mathematical ideas is given in section~3,
allowing the computation of the values of the $p$'s, which turns to
give a new protocol for the WiFi/WiMax networks.  Here again, this
implementation is not necessary to implement our protocol, which is
described in terms of parameters in table~\ref{tab:values}.  Finally,
our protocol is compared with other ones in section 4 where some
numerical results are presented.

\section{Mathematical analysis}

In the following we denote $m$ the number $2^k$.

We denote by $q_n$ the probability that $n$ stations try to emit in
the system. We have necessarily $q_n\geq0$ and $\sum_{n\geq1}q_n=1$.
We introduce $f$, that we will call in the following the
generating function of the distribution of stations, defined by:
 $$f(x)=\sum_{n\geq1}q_nx^n.$$

We try now to characterize the distribution after the transmission
on the first mini-slot. Let $f_1$, respectively $f_0$, be generating functions
for the number of stations still in contention depending, respectively, on
whether or not there was a transmission in the previous slot.
If every station emits a signal with probability $p$, then the
probability that $n$ stations emit is given by:
$$P[\mbox{$n$ stations emit a signal}]=\sum_{i\geq
n}(\,^i_n)q_ip^n(1-p)^{i-n}.$$ Therefore, 
the distribution of the number of stations that emit is
characterized by a function $f_1$ analog of $f$, defined by:
$$f_1(x)=\sum_{i\geq 1}P[\mbox{$i$ stations emit a signal}]x^i.$$
and we deduce logically that:
$$\begin{array}{lcl}
f_1(x)&=&\displaystyle \sum_{n\geq1}\sum_{1\leq i\leq n}(\,_i^n)q_np^i(1-p)^{n-i}x^i\\
&=&\displaystyle \sum_{n\geq1}q_n[(px+1-p)^n-(1-p)^n]\\
&=&f(px+1-p)-f(1-p)
\end{array}.$$
 
Similarly, in the event where no signal has been emitted at the first round,
we can deduce some information on the distribution of the number of stations.
Indeed, the probability that $n$ stations remain silent is $(1-p)^i$.
Therefore if we write:
$$f_0(x)=\sum_{i\geq0}P[\mbox{$i$\small\ stations are present and remain silent}]x^i$$
then we obtain:
$$f_0(x)=\sum_{i\geq0}q_ix^i(1-p)^i=f((1-p)x).$$ 
And we see that at the end of the first round of selection,
the distribution of the whole set of surviving station
can be known by the mathematical function $f_0+f_1$. We can also note that
the distribution in the case where the event $r(0)$ occurs (either 0 or 1)
is given by $f_{r(0)}/f_{r(0)}(1)$.

By extension, if we note $w$ a word in the alphabet $\{0,1\}$ and $w0$
(resp. $w1$) the same word to which the letter ``0'' (resp. ``1'') is
added, and if we note $p_w$ and $f_w$, respectively, the probability
and generating function corresponding to step $w$, then (setting $f_{\emptyset}=f$) the following inductive formulas hold:

$$
\left\{\begin{array}{l}
f_{w1}(x)=f_w(p_wx+1-p_w)-f_w(1-p_w)\\
f_{w0}(x)=f_w((1-p_w)x)
\end{array}\right.
$$

We observe that the probability of the event of the choice
$w=r(1)\dots r(t)$ is $f_w(1)$. Given that the event $w$ occurs, the
distribution of the number of stations is characterized by
$f_{r(1)\dots r(k)}/f_{r(1)\dots r(k)}(1)$.  If we denote by $l(w)$
the length of the word $w$, then the global distribution for all the
event space after $k$ rounds of selection is given by the sum of all
the $f_w$'s that correspond to an event after $k$ rounds, (which is
true if and only if $l(w)=k$), and therefore:
$$g(x)=\sum_{w:l(w)=k}f_w(x).$$
The probability of success $\rho$ of the rounds of selection
is the probability that only one station remains.
It is given by the first term of the integral series of $g$,
that is $g'(0)$. Therefore:
$$\rho=\sum_{w:l(w)=k}f'_w(0).$$
 
In the following we evaluate the value of $f'_w(0)$.

We therefore denote, for any word $w$ in the alphabet $\{0,1\}$, 
the quantity defined inductively by $\delta_{\emptyset}=1$ and
\begin{equation}\label{eq:localstep}
\left\{\begin{array}{l}
\delta_{w0}=(1-p_w)\delta_w\\
\delta_{w1}=p_w\delta_w
\end{array}
\right.
\end{equation}

Then we note $w_a<w_b$ if their corresponding binary values verify this inequality, and, using the convention $y_{\emptyset}=0$, we set:
\begin{equation}\label{eq:cumulstep}
y_w=\sum_{\begin{array}{c}v:l(v)=l(w)\\v<w\end{array}}\delta_v.
\end{equation}

\begin{lem}
$$y_{w0}=y_{w}.$$
\end{lem}
\begin{proof}
It is easy to see that $\delta_{w1}+\delta_{w0}=\delta_w$. 
Therefore  $$
\begin{array}{lcl}
y_{w0}&=&\displaystyle\sum_{\begin{array}{c}v:l(v)=l(w0)\\v<w0\end{array}}\delta_v\\
&=&\displaystyle\sum_{\begin{array}{c}u:l(u)=l(w)\\u<w\end{array}}\delta_{u0}+\delta_{u1}\\
&=&\displaystyle\sum_{\begin{array}{c}u:l(u)=l(w)\\u<w\end{array}}\delta_{u}=y_w
\end{array}$$ 
\end{proof}
\begin{lem}
$f'_w(x)=\delta_wf'(y_w+\delta_wx)$ for all $x\in[0,1]$.
\end{lem}
\begin{proof}
Obviously $\delta_{\emptyset}f'(y_{\emptyset}+\delta_{\emptyset}x)=f'(x)=f'_{\emptyset}$. 
Then we apply another induction on $w$. We suppose that the statement is established for $w$ and show that it is true for $w0$ and $w1$. Indeed:
$$\begin{array}{lcl}
f'_{w0}(x)&=&(1-p_w)f'_w((1-p_w)x)\\
&=&(1-p_w)\delta_wf'(y_w+\delta_w(1-p_w)x)\\
&=&\delta_{w0}f'(y_{w0}+\delta_{w0}x),\\
\end{array}$$
 
moreover, noticing from equation (\ref{eq:cumulstep}) that $y_{w1}=y_{w0}+\delta_{w0}$, we have,
 
$$\begin{array}{lcl}
f'_{w1}(x)&=&p_wf'_w(p_wx+1-p_w),\\
&=&p_w\delta_wf'(y_w+\delta_w(p_wx+1-p_w)),\\
&=&\delta_{w1}f'(y_{w0}+\delta_{w0}+\delta_{w1}x),\\
&=&\delta_{w1}f'(y_{w1}+\delta_{w1}x).\\
\end{array}$$
\end{proof}

And therefore
$$\rho=\sum_{w:l(w)=k}\delta_wf'(y_w).$$

\begin{figure}
\centerline{\psfig{figure=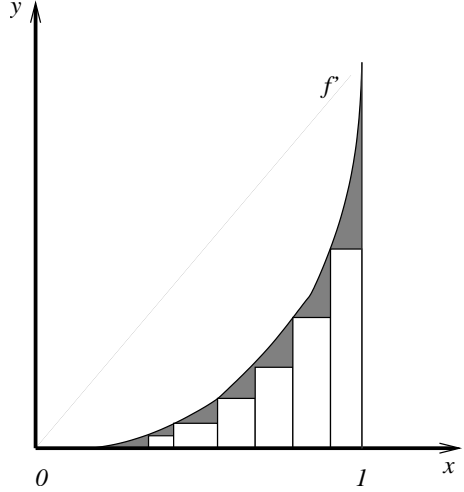,width=6cm}}
\caption{Interpretation of the collision rate in terms of Riemann integral.}
\label{fig:selec-princip}
\end{figure}

This formula exactly says that we aim at approximating
the integral of $f'$ by a Riemann integral. In other words,
if we are given $m-1=2^{k} -1$ real numbers $z_1,\dots,z_{m-1}$ in $]0,1[$,
with $0=z_0<z_1<\dots<z_{m-1}<z_m=1$, then the quantity 
$$\rho=\sum_{i\in\{1,\dots,m\}}(z_i-z_{i-1})f'(z_{i-1})$$ 
is the approximation of the integral of $f'$ by a piece-wise constant function
having $m$ steps. This fact is illustrated by figure~\ref{fig:selec-princip}. 
In this figure we draw the $f'$ function, which integral between $0$ and $1$ 
exactly equals $1$ (since $f(1)-f(0)=1$). Points have been chosen to approximate this integral by a lower-bound piece-wise constant function. The rate of collision
of our protocol will be exactly equal to the area in gray on the figure,
which is also the approximation default. Therefore, if we have the best values
of $z_0,\dots,z_m$ for this integral, it is sufficient to set 
$y_w=z_{\#(w)}$, where $\#(w)$ is the numerical binary value that is represented by $w$. The reader can verify that this is obtained by setting:
$$p_w=\frac{z_{\#(w)2^{k-l(w)}+2^{k-l(w)}}-z_{\#(w)2^{k-l(w)}+2^{k-l(w)-1}}}{z_{\#(w)2^{k-l(w)}+2^{k-l(w)}}-z_{\#(w)2^{k-l(w)}}}.$$

\begin{prop}
For all protocol of selection governed by a series of selective rounds
as indicated in figure~\ref{fig:802.11c}, we can associate a series of
$m+1$ real numbers $z_0,\dots,z_m$ in $[0,1]$, with $0=z_0<z_1<\dots<z_{m-1}<z_m=1$, such as the probability of success (non-collision) of the protocol is given by:
\begin{equation}\label{eq:rho}
\rho=\sum_{i\in\{1,\dots,m\}}(z_i-z_{i-1})f'(z_{i-1}).
\end{equation}

In this case, the probabilities chosen to operate the different rounds of selection are given by:
\begin{equation}\label{eq:pw}
p_w=\frac{z_{\#(w)2^{k-l(w)}+2^{k-l(w)}}-z_{\#(w)2^{k-l(w)}+2^{k-l(w)-1}}}{z_{\#(w)2^{k-l(w)}+2^{k-l(w)}}-z_{\#(w)2^{k-l(w)}}}.
\end{equation}

Inversely, to all family of real numbers $z_0,\dots,z_m$ verifying
$$0=z_0<z_1<\dots<z_{m-1}<z_m=1,$$ we can associate a protocol of
selection whose probability of success and probabilities are given by
equations~(\ref{eq:rho}) and (\ref{eq:pw}).
\end{prop}

So we are now left with the problem of finding optimal values for $z$.
Analyzing a little further the value of $\rho$, we obtain the formula:
$$\begin{array}{lll}
1-\rho&=&\displaystyle\int_0^1f'(t)dt-\sum_{i\in\{1,\dots,m\}}(z_i-z_{i-1})f'(z_{i-1})\\
&=&\displaystyle\sum_{i\in\{1,\dots,m\}}f(z_i)-f(z_{i-1})-(z_i-z_{i-1})f'(z_{i-1}).
\end{array}$$

Three lemmas will allow us to analyze it.

\begin{lem} \label{lem:bound_error}
Let $\hbar$ be the piece-wise constant function defined by
\begin{equation}\label{eq:hbar}
\hbar:x\mapsto\frac1{m(z_i-z_{i-1})}\mbox{ for }x\in[z_{i-1};z_i[.
\end{equation}
We have $\int^0_1\hbar(t)dt=1$  and  
$$\begin{array}{l}
\displaystyle\frac{1}{2m}\int_0^1\frac{f''(t)}{\hbar(t)}dt\\
=\displaystyle\sum_{i\in\{1,\dots,m\}}\int^{z_i}_{z_{i-1}}\frac{z_i-z_{i-1}}2f''(t)dt\\
=\displaystyle\sum_{i\in\{1,\dots,m\}}\frac{(z_i-z_{i-1})(f'(z_i)-f'(z_{i-1}))}2.
\end{array}$$

Moreover,
$$1-\rho-\frac1{2m}\int^1_0\frac{f''(t)}{\hbar(t)}dt\geq-\frac1{12}\sum_{i\in\{1,\dots,m\}}(z_i-z_{i-1})^3f'''(z_i)$$
and
$$1-\rho-\frac1{2m}\int^1_0\frac{f''(t)}{\hbar(t)}dt\leq-\frac1{12}\sum_{i\in\{1,\dots,m\}}(z_i-z_{i-1})^3f'''(z_{i-1}).$$
\end{lem}

\begin{proof}
$$\begin{array}{l}
\displaystyle
1-\rho-\frac1{2m}\int^1_0\frac{f''(t)}{\hbar(t)}dt\\
\displaystyle=\sum_{i\in\{1,\dots,m\}}f(z_i)-f(z_{i-1})-(z_i-z_{i-1})f'(z_{i-1})\\
\qquad\qquad\qquad-\frac{(z_i-z_{i-1})(f'(z_i)-f'(z_{i-1}))}2\\
\displaystyle=\sum_{i\in\{1,\dots,m\}}f(z_i)-f(z_{i-1})\\
\qquad\qquad\qquad-(z_i-z_{i-1})\frac{f'(z_i)+f'(z_{i-1})}2.\\
\end{array}
$$      
 
But $f$ is an harmonic function with positive coefficients, and radius
of convergence from 0 at least 1, therefore:
 $$
\begin{array}{l}
\displaystyle f(z_i)=\sum_{j\geq0}\frac{(z_i-z_{i-1})^j}{j!}f^{(j)}(z_{i-1})\\
\displaystyle f'(z_i)=\sum_{j\geq1}\frac{(z_i-z_{i-1})^{j-1}}{(j-1)!}f^{(j)}(z_{i-1})\\
\end{array}
$$
and we have
$$\begin{array}{l}
f(z_i)-f(z_{i-1})-\frac{z_i-z_{i-1}}2(f'(z_i)+f'(z_{i-1}))\\
=(z_i-z_{i-1})^3{\displaystyle\sum_{j\geq3}}(z_i-z_{i-1})^{j-3}\left[\frac1{j!}-\frac1{2(j-1)!}\right]\times\\\qquad f^{(j)}(z_{i-1}),\\
=-(z_i-z_{i-1})^3{\displaystyle\sum_{j\geq3}}(z_i-z_{i-1})^{j-3}\frac{1}{(j-1)!}\left[\frac12-\frac1{j}\right]\times\\\qquad f^{(j)}(z_{i-1}).
\end{array}$$
We note that all the derivatives of $f$ are positive, and therefore all the term of this series are non-positive. Hence the second inequality. Moreover,
 
$$\begin{array}{l}
\displaystyle\sum_{j\geq3}(z_i-z_{i-1})^{j-3}\left[\frac1{2(j-1)!}-\frac1{j!}\right]f^{(j)}(z_{i-1})\\
\displaystyle\quad=\sum_{j\geq3}(z_i-z_{i-1})^{j-3}\frac{j-2}{2(j!)}f^{(j)}(z_{i-1})\\
\displaystyle\quad\leq\sum_{j\geq3}(z_i-z_{i-1})^{j-3}\frac{1}{12(j-3)!}f^{(j)}(z_{i-1})\\
\displaystyle\quad\leq\frac1{12}f'''(z_i)
\end{array}$$
\end{proof}

\begin{lem} \label{lem:bound_square}
 Suppose $f''(0)>0$. Let the real numbers $z_0,\dots,z_m$ in $[0,1]$, with $0=z_0<z_1<\dots<z_{m-1}<z_m=1$, that achieve the maximum value of $\rho=\sum_{i\in\{1,\dots,m\}}(z_i-z_{i-1})f'(z_{i-1})$. Then we have:
$$\sum_{i\in\{1,\dots,m\}}(z_i-z_{i-1})^2\leq\frac{2}{mf''(0)},$$and$$\forall i\in\{1,\dots,m\}\quad z_i-z_{i-1}\leq\sqrt{\frac{2}{f''(0)}}\frac{1}{\sqrt{m}}.$$  
\end{lem}

\begin{proof}
We have:
$$1-\rho=\sum_{i\in\{1,\dots,m\}}f(z_i)-f(z_{i-1})-(z_i-z_{i-1})f'(z_{i-1}),$$
 
and by Taylor expansion, there exists $b_{i-1}$ in the interval
$]z_{i-1};z_i[$ such that
 
$$f(z_i)-f(z_{i-1})=$$
$$(z_i-z_{i-1})f'(z_{i-1})+\frac{(z_i-z_{i-1})^2}{2}f''(b_{i-1}).$$

Therefore $\displaystyle 1-\rho=\sum_{i\in\{1,\dots,m\}}\frac{(z_i-z_{i-1})^2}{2}f''(b_{i-1}).$

Since $\rho$ is a maximum value for all the choices of $z_i$,
necessarily $1-\rho\leq1/m$ (indeed, if we take $z_i=i/m$, we obtain a
solution verifying $1-\rho\leq1/m$). Therefore $z_i-z_{i-1}$ tends to
$0$ when $m$ tends to infinity.  The previous inequality taken term by
term gives $z_i-z_{i-1}\leq\sqrt{\frac{2}{f''(0)}}\frac1{\sqrt{m}}$,
and since all the $f''(b_i)$ are bounded below by $f''(0)$, we obtain
$$\sum_{i\in\{1,\dots,m\}}(z_i-z_{i-1})^2\leq\frac{2}{mf''(0)}.$$
\end{proof}

\begin{lem} \label{lem:mini_h}
If $f''$ is piece-wise continuous,
 the minimum of the value $\int_0^1\frac{f''(t)}{h(t)}dt$ on the functions $h$ piece-wise continuous and verifying $\int_0^1h(t)dt=1$ is obtained
by
$h^*:x\mapsto\frac{\sqrt{f''(x)}}{\int_0^1\sqrt{f''(t)}dt}$ and therefore
equals $\left(\int_0^1\sqrt{f''(t)}dt\right)^2$.
\end{lem}
The proof of lemma~\ref{lem:mini_h} is easily obtained if
$f''$ is a piecewise constant-function, and we use uniform convergence of
piecewise constant-functions to piece-wise continuous functions.

\begin{prop}\label{prop:contprob}
 If $f''(0)>0$, whatever the series of $m$ values of $z_.$ used, we have:

$$1-\rho\geq\frac{\left(\int_0^1\sqrt{f''(t)}dt\right)^2}{2m}-\frac{f''(1)}{3\sqrt{2}f''(0)^{3/2}}\frac1{m^{3/2}}.$$
\end{prop}

\begin{proof}
By lemma (\ref{lem:bound_error}) we have:
$$1-\rho\geq\frac1{2m}\int^1_0\frac{f''(t)}{\hbar(t)}dt-\frac1{12}\sum_{i\in\{1,\dots,m\}}(z_i-z_{i-1})^3f'''(z_i).$$
Then applying lemma (\ref{lem:mini_h}) for $\hbar=h$ gives
$$1-\rho\geq\frac1{2m}\int^1_0\frac{f''(t)}{\hbar(t)}dt-\frac1{12}\sum_{i\in\{1,\dots,m\}}(z_i-z_{i-1})^3f'''(z_i).$$
Then the second inequality of lemma~(\ref{lem:bound_square}) gives:
$$
\begin{array}{l}
\displaystyle 1-\rho\geq\frac1{2m}\int^1_0\frac{f''(t)}{\hbar(t)}dt\\
\qquad\displaystyle-\frac1{12\sqrt{m}}\sqrt{\frac{2}{f''(0)}}\sum_{i\in\{1,\dots,m\}}(z_i-z_{i-1})^2f'''(z_i).
\end{array}
$$
Since $f'''(z_i)\leq f'''(0)$, it gives:
$$
\begin{array}{l}
\displaystyle 1-\rho\geq\frac1{2m}\int^1_0\frac{f''(t)}{\hbar(t)}dt\\
\qquad\displaystyle-\frac{f'''(0)}{12\sqrt{m}}\sqrt{\frac{2}{f''(0)}}\sum_{i\in\{1,\dots,m\}}(z_i-z_{i-1})^2.
\end{array}
$$
Finally the first inequality of lemma~(\ref{lem:bound_square}) gives:
$$
1-\rho\geq\frac1{2m}\int^1_0\frac{f''(t)}{\hbar(t)}dt
-\frac{f'''(0)}{12\sqrt{m}}\sqrt{\frac{2}{f''(0)}}\frac{2}{mf''(0)}.
$$
Hence the result.
\end{proof}

This result clearly shows what is the maximum
efficiency of the protocol whatever the chosen values of probabilities,
and therefore the maximum we can asymptotically expect, that is
$$1-\rho\approx\frac{\left(\int_0^1\sqrt{f''(t)}dt\right)^2}{2m}.$$
The strategy of approaching $h$ by a piecewise constant-function
therefore gives a result asymptotically optimal.

We note also that as the number of round increases, the collision
rates systematically decreases, and this factor of division converges
to 2 as the number of rounds increases. It means that we can reduce
the collision rate by an arbitrary low level by using a reasonably
small number of additional rounds. For applications that suffer from
retransmissions (and jitter) this property can have a considerable
impact. 

However, further experiments showed, that in the 802.11b framework
where the jitter was not important and the throughput was to be optimized,
the good number of mini-slots to use was 6.

Some more results on the approximation of a function by Riemann
integrals are worth to be noted. We are interested in the following
problem:

\begin{prob}
Let $\varphi$ a non-decreasing function from $[0;1]$ to $[0;1]$, with
$\varphi(0)=0$ and $\varphi(1)=1$. We are looking for a piecewise
constant function $\psi$, with $m+1$ pieces, such that
$$
\left\{
\begin{array}{l}
\forall z\in[0;1]\quad\psi(z)\leq\varphi(z),\\
\displaystyle\int_0^1\left(\varphi(z)-\psi(z)\right)dz\mbox{ is minimum}.
\end{array}
\right.
$$
\end{prob}
Clearly, this problem is the optimization formulation of the preceding issue
when, for $z\in[0;1]$, $\varphi(z)=f'(z)/f'(1)$.
 
\begin{prop}
If $z\mapsto z\varphi(z)$ is convex, $m=1$, and $\varphi$ continuously
derivable, then the minimum for $g$ is reached with a step $z$
verifying $(1-z)\varphi'(z)=\varphi(z)$.
\end{prop}

\begin{proof}
If the step is $z\in[0,1]$, then
$$\int_0^1(\varphi(t)-\psi(t))dt=\int_0^1\varphi(t)dt-(1-z)\varphi(z).$$
Note that necessarily there is some $z^*\in]0;1[$ such that
$\varphi(z^*)>0$ and this particular $z^*$ does better than $z=0$
or $z=1$. The best $z$ necessarily then verifies $\frac{d}{dz}(1-z)\varphi(z)=0$. 
\end{proof}

\begin{prop}
For a fixed $m$, for any $\varphi$ function, there is a $\psi$
function
that reaches the minimum. This minimum will be noted $c_m$.
\end{prop}

\begin{proof}L
Set $$c_m=\inf_{\mbox{\begin{tabular}{l}$\psi$ piece wise constant\\with
  $m+1$ pieces,\\ and
  $\psi\leq\varphi$\end{tabular}}}(\varphi(t)-\psi(t))dt.$$ 
Let $\psi_p$ be a series of piece-wise constant functions, with $m+1$
pieces, such that
$$\lim_{p\rightarrow\infty}\int_O^1(\varphi(t)-\psi_p(t))dt=c_m.$$

Let $z^{(1)}_p\leq\dots\leq z^{(m)}_p$ be the points where $g_p$ is not continuous. Since $[0;1]$ is compact, let us extract a converging
  series $z^{(1)}_{\sigma_1(p)}$ (that is, $\sigma_1$ is an increasing
  function from $\N$ to $\N$ such that the series
  $z^{(1)}_{\sigma_1(p)}$, $p\in\N$ is converging), and $\sigma_2$
  such that $z^{(2)}_{\sigma_2(\sigma_1(p))}$ is converging, and so on
  to $\sigma_m$ such that $z^{(m)}_{\sigma_m o\dots o\sigma_1(p)}$ is
  converging.  Setting $\sigma=\sigma_m o\dots o\sigma_1$ we have that
  $\sigma$ is increasing from $\N$ to $\N$ and for each
  $i\in\{1,\dots,m\}$, $z^{(i)}_{\sigma(p)}$, $p\in\N$ is converging.

Let us then note $z^{(i)}_*=\lim_{p\rightarrow\infty}z^{(i)}_{\sigma(p)}$,
for $i\in\{1,\dots,m\}$, $z^{(0)}_*=0$, $z^{(m+1)}_*=1$,
and $\psi^*$ such that for $i\in\{1,\dots,m+1\}$, and
$z\in[z^{(i-1)}_*,z^{(i)}_*[$, $\psi^*(z)=\varphi(z^{(i-1)}_*)$.

Let $\varepsilon>0$ be a real number. There is a $\eta>0$ such that
for all $i\in\{1,\dots,m\}$, $|z-z^{(i)}_*|<\eta$ implies
$|\varphi(z)-\varphi(z^{(i)}_*)|<\varepsilon$. If $\eta>\varepsilon$, set
$\eta=\varepsilon$.  Then there is a $P\in\N$ such that $p\geq P$
implies $|z^{(i)}_*-z^{(i)}_{\sigma(p)}|\leq\eta$.

We see that for each $i\in\{1,\dots,m\}$,
$$\psi_*(z^{(i)}_*)=\varphi(z^{(i)}_*)\geq\varphi(z^{(i)}_p)-\varepsilon=\psi_p(z^{(i)}_p)-\varepsilon$$
		     for $p\geq P$, and therefore
$$\int_0^1 \left(\psi_*(t)-\psi_p(t)\right)dt\geq
-\varepsilon-m\eta\geq-(m+1)\varepsilon.$$
This is true for all $\varepsilon>0$, and so $\int_0^1\psi_*(t)dt\geq c_m.$
\end{proof}

\begin{prop}
Let $A$ the function from $\R^m$ to $\R$ given by
$$A(z_1,\dots,z_m)=$$
$$\sum_{i=2}^{i=m}
(z_i-z_{i-1})\varphi(z_{i-1})+(1-z_{m})\varphi(z_m).$$

Then $$\max_{(z_1,\dots,z_m)\in[0;1]^m} A(z_1,\dots,z_m)=c_m.$$
\end{prop}

\begin{proof}
It suffices to show that for the maximum we have $z_1\leq\dots\leq
z_m$. Note that if $\alpha<\beta$, then
$$(\beta-\alpha)\varphi(\alpha)+(1-\beta)\varphi(\beta)>(\alpha-\beta)\varphi(\beta)+(1-\alpha)\varphi(\alpha).$$
It follows that
$$A(z_1,\dots,z_{i-1},\alpha,\beta,z_{i+1},\dots,z_m)\geq$$
$$A(z_1,\dots,z_{i-1},\beta,\alpha,z_{i+1},\dots,z_m).$$ Therefore,
ordering the arguments of $A$ maximizes its results (use for instance
bubble sort).
\end{proof}

Those results open tools for promising optimization.

\section{Practical implementation}

\begin{figure}
\centerline{\psfig{figure=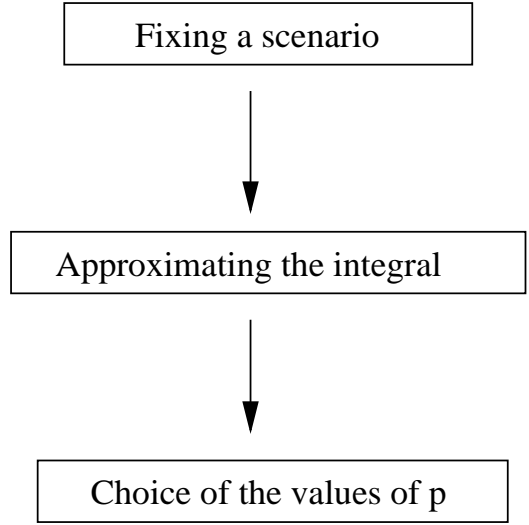,width=7cm}}
\caption{Strategies of optimization of the values of $p$}
\label{fig:strateg-802.11c}
\end{figure}

In figure~\ref{fig:strateg-802.11c}, we show the basic principles of
an optimization based on our mathematical analysis. A preliminary step
consists in fixing a scenario, that is the probabilities that a given
number of stations appears.  For instance we set as previously
$$P[\mbox{Number of emitting stations} = n]=q_n.$$ 
And we consider 
 $$f(x)=\sum_{n\geq1}q_nx^n.$$ 
Then we have an $\hat{h}$ function defined by
$\hat{h}(x)=\sqrt{f''(x)}$ ($\hat{h}$ is the equivalent
of $h^*$ in the previous section, but without the normalization
$\int_0^1 h^*(t)dt=1$).
And we take a number $M$ largely greater than $m=2^k$
and we compute $H$ as follows
$$\left\{\begin{array}{l}
H(0)=0\\
H(i+1)=H(i)+\hat{h}\left(\frac{i+1/2}{M}\right)\\
\qquad\mbox{ for }i\in\{0,\dots,M-1\}\\
\end{array}\right.$$
and we define $z_j$ for $j\in\{0,\dots,m\}$  by
$$\left\{
\begin{array}{l}
z_0=0,\\
z_j=\frac1M\min\left\{i:\frac{H(i)}{H(M)}\geq\frac{j}{m}\right\}\\
\qquad\mbox{ for }j\in\{1,\dots,m-1\},\\
z_m=1.
\end{array}
\right.
$$

Finally we set
$$p_w=\frac{z_{\#(w)2^{k-l(w)}+2^{k-l(w)-1}}-z_{\#(w)2^{k-l(w)}+2^{k-l(w)}}}{z_{\#(w)2^{k-l(w)}}-z_{\#(w)2^{k-l(w)}+2^{k-l(w)}}}$$
where $w$ is a word in the $\{0,1\}$ alphabet, $\#(w)$ represents the
numerical binary value denoted by $w$ and $l(w)$ the length of the
word $w$.

\section{Numerical results}

In the first part of this section, we compare the collision rate of our
method to that of CONTI \cite{AC05}. We show that for some parameter
our collision rate is always favorable, and therefore systematically
results in a better performance to CONTI.  Along with the native
802.11b protocol \cite{IEEEnorm}, we compare to two high performing
protocols, the Idle Sense one \cite{HRG+05} and the additive
congestion window in\-crea\-se~/\-de\-crea\-se protocol
\cite{Galtier04a}. Finally we concentrate on fairness issues for these
different schemes, based on the Jain index \cite{JCH84}.

\subsection{A tuning of the probabilities}

\begin{table}
\centerline{
\begin{tabular}{|l|cccccc|}\hline
$l(w)$&0&1&2&3&4&5\\
$p_w$&0.07&0.2&0.25&0.33&0.4&0.5\\\hline
\end{tabular}
}
\caption{Values taken by CONTI.}
\label{tab:conti}
\end{table}

\begin{figure}
\centerline{\psfig{figure=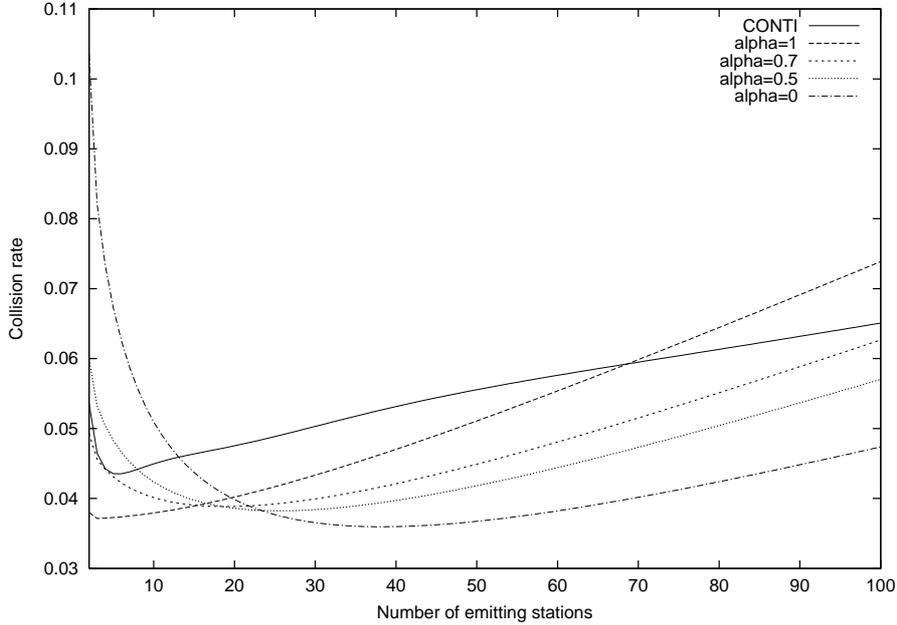,width=12cm}}
\caption{Various collision rates obtained with our algorithm
compared to CONTI.}
\label{fig:rates}
\end{figure}

An essential step is to set the values of $q_n$. One idea is to
set a favorite interval of operation, say $\{2,\dots,N\}$,
and fix, for $n\in\{2,\dots,N\}$, and some $\alpha\in[0,1]$,
\begin{equation}\label{eq:setq}
q_n=\frac{n^{-\alpha}}{\sum_{i=2}^{i=N}i^{-\alpha}}.
\end{equation}
This distribution allows to take into account in a balanced way loaded
or non loaded networks. In figure~\ref{fig:rates}, we show different
probability curves obtained for $N=100$, and various values of
$\alpha$.

In order to see the advantage of our optimization techniques, we
compare our results to that performed by CONTI \cite{AC05}. In our
context, one can simply view CONTI as a special assignment of the
values of $p_w$ that only depends on the length $w$. For completeness,
we recall the values taken in table~\ref{tab:conti}.

Globally, we can see that
these functions perform remarkably well compared to CONTI. Whereas
the latter has a collision rate between 4.5 and 6.5\%, with a maximum for
the 4.5\% rate, our algorithms fall often under 4\%.  The value
$\alpha=0$ allows to give equal weights to all the events. In
practice, we see that this global optimization tends to
pay more attention to the cases where more stations are present (50 to
100) at the expense of more collisions when two to five stations are
present in the system. On the contrary $\alpha=1$ performs well when a
small number of stations are present at the expense of lesser
performance over 60 stations. A good compromise seems to be
$\alpha=0.7$ which varies between 3.9\% and 6.3\% with an average
improvement to CONTI of 13.9\% in collision rate. For the cases
$\alpha=0$ and $\alpha=0.5$, the improvement - although in some cases
negative - is on average even better, respectively 21.1\% and 17.8\%.
In the following we will set $\alpha=0.7.$ For sake of completeness,
we give in figure~\ref{tab:values} our probability values
so that the reader can replicate our experiments without
further considerations on choosing $\alpha$.

\subsection{Comparative bandwidth}

\begin{table}
$$
\begin{array}{|ll|ll|}\hline
&&p_{00000}&0.470679\\
p_{}&0.0628357&p_{00001}&0.471427\\
p_{0}&0.166808&p_{00010}&0.472147\\
p_{1}&0.305488&p_{00011}&0.472882\\
p_{00}&0.295586&p_{00100}&0.473527\\
p_{01}&0.328258&p_{00101}&0.474214\\
p_{10}&0.375175&p_{00110}&0.474668\\
p_{11}&0.423688&p_{00111}&0.475313\\
p_{000}&0.388521&p_{01000}&0.476041\\
p_{001}&0.398651&p_{01001}&0.476681\\
p_{010}&0.407585&p_{01010}&0.477124\\
p_{011}&0.416295&p_{01011}&0.477647\\
p_{100}&0.429211&p_{01100}&0.477976\\
p_{101}&0.444548&p_{01101}&0.478795\\
p_{110}&0.457931&p_{01110}&0.478203\\
p_{111}&0.465291&p_{01111}&0.48056\\
p_{0000}&0.442201&p_{10000}&0.479927\\
p_{0001}&0.444984&p_{10001}&0.480932\\
p_{0010}&0.447669&p_{10010}&0.481663\\
p_{0011}&0.450083&p_{10011}&0.48324\\
p_{0100}&0.452293&p_{10100}&0.484177\\
p_{0101}&0.454545&p_{10101}&0.485714\\
p_{0110}&0.456444&p_{10110}&0.486056\\
p_{0111}&0.459286&p_{10111}&0.486726\\
p_{1000}&0.462745&p_{11000}&0.490291\\
p_{1001}&0.466754&p_{11001}&0.491979\\
p_{1010}&0.469799&p_{11010}&0.491329\\
p_{1011}&0.473795&p_{11011}&0.490566\\
p_{1100}&0.475827&p_{11100}&0.489796\\
p_{1101}&0.478916&p_{11101}&0.492754\\
p_{1110}&0.484211&p_{11110}&0.492188\\
p_{1111}&0.483871&p_{11111}&0.491667\\
\hline
\end{array}
$$
\caption{Values obtained for the $p_\cdot$'s for $\alpha=0.7$
and $N=100$.}
\label{tab:values}
\end{table}

\begin{figure}
\centerline{\psfig{figure=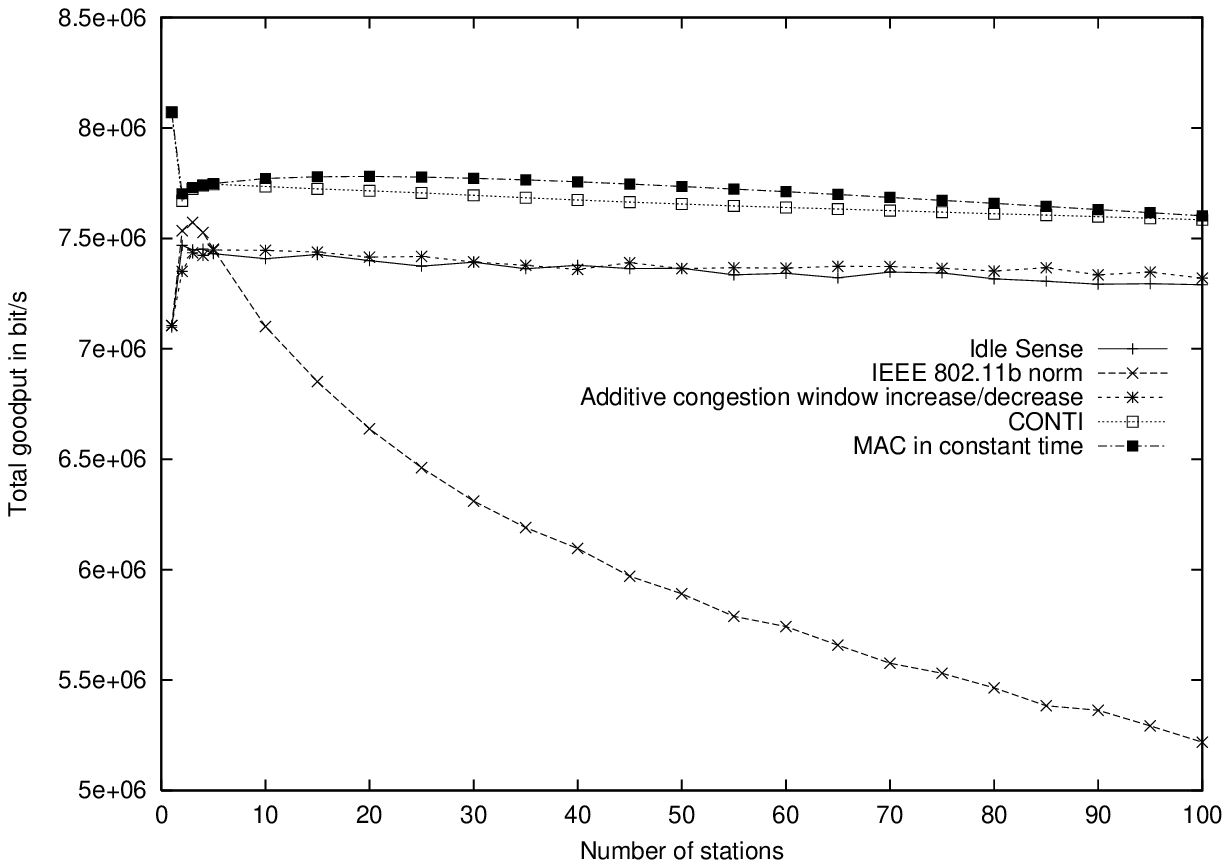,width=12cm}}
\caption{Comparative total throughput for different protocols.}
\label{fig:throughput}
\end{figure}

We set the general parameters as follows, according to the IEEE
802.11b norm. The SIFS and DIFS times are set to $10\mu s$ and $50\mu
s$ respectively.  The time-slot interval for CRP is set to $20\mu
s$. The size of the payload of a packet is set to 1500 bytes. A packet
(either regular or ACK) contains a physical header of $96\mu s$. On
top of that, the MAC head and tail represent in all 19 bytes in a
regular packet, and 14 bytes in an ACK one, that are transmitted at
the maximum speed, that is 11Mbit/s. We now further describe the
specificity of each protocol.

\begin{itemize}
\item{\bf The 802.11b norm \cite{IEEEnorm}.} Each station has a $CW$ parameter.
At the beginning, a station chooses a $\kappa$ - called back-off counter -
randomly in the interval $\{0,\dots,CW-1\}$. If $\kappa=0$, the transmission
begins immediately. Otherwise, if an empty time-slot is observed, 
$\kappa$ is decreased by one. At the end of a transmission,
the value of $CW$ itself is updated to $CWMin$ if the transmission
was successful, and to $\min(CWMax,2*CW)$ if a collision occurred.
We have set as in the norm $CWMin=32$ and $CWMax=1024$.
\item{\bf The Idle Sense method \cite{HRG+05}.}
At the end of a transmission, successful or not, the terminal
stores the number of idle time-slots before its transmission.
After 5 transmissions, the terminal computes the average waited
time-slots. If this number is inferior to $5.68$, the congestion
window is updated by
$$CW=\min(CWMax,CW*1.2).$$
Otherwise, the new $CW$ is given by:
$$CW=\max(CWMin,2*CW/(2+1e-3*CW)).$$
\item{\bf The additive congestion window increase/de\-crea\-se \cite{Galtier04a}.}
At the end of an unsuccessful transmission, $CW$ is set
to $\min(CWmax,CW+32)$. If the transmission is successful,
the station flips a biased coin, and with probability $0.1809$
updates $CW$ by $$CW:=\max(CWmin,CW-32)$$ and otherwise does not change $CW$.
\item{\bf The CONTI method \cite{AC05}.} At each step a CRP of six time-slots is
applied with the probabilities given by table~\ref{tab:conti}.
The surviving stations transmit.
\item{\bf Our method - MAC in fixed congestion window.} We apply a CRP of six time-slots.
We have used the probabilities of table~\ref{tab:values}.
\end{itemize}
Our results are presented in figure~\ref{fig:throughput}. In this
figure, we plot for various numbers of stations the total throughput
observed in the system. We clearly see that all the proposed methods
improve significantly the original IEEE 802.11b mechanism. The methods
based on adaptive tuning of the congestion window, namely
\cite{HRG+05,Galtier04a}, achieve quite close performances. The CONTI
method performs very well. Our method gives the best
performance in all cases, and has a total improvement as far as 31.4\%
for 100 stations to the original norm.

\subsection{Fairness considerations}

\begin{figure}
\centerline{\psfig{figure=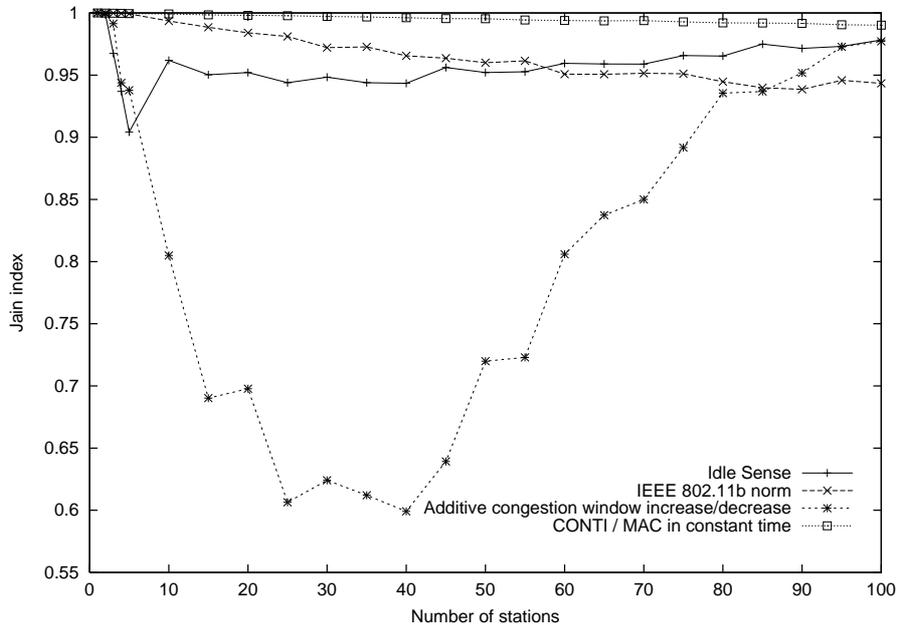,width=12cm}}
\caption{Comparative Jain index for different protocols.}
\label{fig:jain}
\end{figure}

In this part we take into consideration the fairness issues. For each
of the experiments, we have observed a series of 10000 successful
transmission, and assigned to each station $i$ the number $x_i$ of
packets it managed to transmit. In order to evaluate the fairness, we
use the Jain index \cite{JCH84}, defined by:
$$Index=\frac{\left(\sum x_i\right)^2}{n\sum x_i^2}.$$ This index is
always between $0$ and $1$, and closer to $1$ if the system is more
fair. Our results are given in figure~\ref{fig:jain}.  Note that our
method is equivalent to the CONTI method from the fairness point of
view. The results plotted are averages obtained after a series of 10
tests.

The results show different behaviors. We observe as in \cite{HRG+05}
that slow congestion window methods tend to generate some unfairness.
We also notice that new method hardly improve the quality of the
original IEEE 802.11b norm.  Note, anyway, that our method achieves
the best fairness performance.

\section{Conclusion}
In this paper we have demonstrated the efficiency of MAC protocols
with constant congestion window size in the wireless context. We have
determined their limits in terms of avoidance of collision, and shown
that they perform very well in terms of fairness. The tuning that we
propose achieves the best throughput performance in the 802.11b
framework to our knowledge.  This advocate for a more extensive use of
these methods, and the building of devices including this new access
control mode. This is not necessarily a simple task, since the
proposed scheme is not compatible with the previous ones excepted
CONTI, but is a promising way to achieve better wireless networks.

\begin{table*}
\centerline{
\begin{tabular}{ll}
$k$&Number of rounds of selection.\\
$m$&$2^k$.\\
$r(t)$&Try-bit at the $t^{\mbox{th}}$ round of selection, $t\in\{1,\dots,k\}$.\\
$R(t)$&Local try-bit (at a station) at the $t^{\mbox{th}}$ round of selection, $t\in\{1,\dots,k\}$.\\
$q_n$&Probability that $n$ station try to emit.\\
$w$&Word in the alphabet $\{0,1\}$.\\
$l(w)$&Length of the word $w$.\\
$\#(w)$&Binary value represented by $w$\\
$p$&Probability that a station emits a signal at the first round of selection.\\
$p_w$&Probability that a non-eliminated station emits a signal at the round $l(w)+1$,\\
&given that the preceding try-bits where $r(1),\dots,r(l(w))=w$.\\
$f$&Generating function of $q_.$, see equation (\ref{eq:deff}).\\
$f_w$&Generating function of the number of non-eliminated stations,
in the event $w$.\\
$g$&Generating function of the number of non-eliminated stations after $k$ rounds.\\
$\rho$&Success rate (as opposed to the collision rate).\\
$\delta_w$&Local step, see equation~(\ref{eq:localstep}).\\
$y_w$&Cumulative step, see equation~(\ref{eq:cumulstep}).\\
$z_i$&Riemann steps for $f'$ in $[0,1]$, $i\in\{0,\dots,m\}$.\\
$\varphi$&Continuous function in $[0;1]$ to be approximated.\\
$\psi$&Piece-wise constant function that approximate $\varphi$.\\
$c_m$&Best approximation gap for the approximation of $\varphi$ with $m$ pieces.\\
$\sigma$&Increasing function from $\N$ to $\N$ that emphasizes extraction.\\
$A$&Function of $\R^{m+1}$ to $\R$ that reaches approximation for $\varphi$.\\ 
$\hbar$&Density step function defined after $z_\cdot$, see equation (\ref{eq:hbar}).\\
$h^*$&Density function that minimizes $\int_0^1f''(t)/h(t)dt$, see proposition~\ref{prop:contprob}.\\
$\hat{h}$&Density function taken for the algorithmic choices of $z_\cdot$, $\hat{h}(x)=\sqrt{f''(x)}$.\\
$M$&Large number compared to $m$.\\
$N$&Maximum number of foreseen stations.\\
$\alpha$&Parameter to set the values of $q_\cdot$, see equation~(\ref{eq:setq}).\\
$x_i$&In the experiments, amount of packets that an individual station has emitted.\\
\end{tabular}
}
\caption{Notations used in this paper.}
\end{table*}

\small

\bibliography{refbib}

\end{document}